# Spreadsheet Refactoring

Patrick O'Beirne

Systems Modelling Ltd

Tara Hill, Gorey, Co. Wexford, Ireland

Tel +353-53-942-2294

pob@sysmod.com

## ABSTRACT

*Refactoring is a change made to the internal structure of software to make it easier to understand and cheaper to modify without changing its observable behaviour.*

*A database refactoring is a small change to the database schema which improves its design without changing its semantics.*

*The paper shall present example 'spreadsheet refactorings', derived from the above and taking into account the unique characteristics of spreadsheet formulas and VBA code. The techniques are constrained by the tightly coupled data and code in spreadsheets.*

*Keywords:  spreadsheets, spreadsheet errors, quality, end-user computing, software development, agile.*

## Refactoring out debt

*Refactoring*[1] is a change made to the internal structure of software to make it easier to understand and cheaper to modify without changing its observable behaviour.

It is the paying off of technical debt. *Technical Debt*[2] is incurred whenever a shortcut is taken that gains functionality at the expense of maintainability. Whenever you insert a quick fix or workaround, you create an extra dependency that must be maintained; but its very manner of creation increases the likelihood of its being overlooked and the risk of error being introduced later. Without a full comprehension of the design of the code, it is very easy to create code that does the same or similar thing in more than one place. This creates opportunities to cause problems by later changing in one place but not another.

Gary Short [3]: "Of course, technical debt is not always a bad thing. A little technical debt can speed the development process. Much like a business might go into financial debt to enable it to purchase the raw materials in order to manufacture the widgets for that big order they've just secured. In the same way, a development team might incur a little technical debt in order to meet an important milestone. Technical debt, like financial debt, only becomes a problem when it is left to get out of control and gather crippling interest."

Left unchecked, the code can become a software Hydra which sprouts two new bugs for every one you attempt to correct. It may take another person a week to make a change to your code which they could have done in an hour if they had understood it correctly. A little time spent refactoring can make the code better communicate its purpose. Programming in this mode is all about saying exactly what you mean (Martin Fowler [4] of the "Gang Of Four"). Kent Beck says about himself, "I'm not a great programmer; I'm just a good programmer with great habits." Refactoring helps you be much more effective at writing robust code.

Refactoring is not bug fixing; nor is it enhancement. It is simply an investment now to reduce later time and effort and risk. It is more than simply the replacement of slow, amateurish formulas with more efficient ones; refactoring is generally in the context of skilled development. However, it is always possible to learn better practices and improve one's ability to implement well-crafted solutions. Therefore in this paper some frequently seen poor designs will be picked on for improvement.

The whole point of having a good design is to allow rapid development. Without a good design, you can progress quickly for a while, but soon the poor design starts to slow you down. You spend time finding and fixing bugs instead of adding new function. Changes take longer as you try to understand the system and find the duplicate code. New features need more coding as you patch over a patch that patches a patch on the original code base.

To adapt Roedy Green the author of "How to Write Unmaintainable Code" [5] "To foil the spreadsheet maintainer, you have to understand how he thinks. He has your giant workbook. He has no time to read it all, much less understand it. He wants to rapidly find the place to make his change, make it and get out and have no unexpected side effects from the change. You want to make it as hard as possible for him to find the code he is looking for. But even more important, you want to make it as awkward as possible for him to safely ignore anything. "

*If we look at our work immediately after completing it, we are still too involved; if too long afterwards, we cannot pick up the thread again.* (Blaise Pascal)

## *Three groups of refactorings*

1. **For worksheets**
   a. Impose Style conventions
   b. Follow Industry Conventions
   c. Make it easy to see non-obvious content
   d. Duplicated code
   e. Overlong formula
   f. Simplify formulas
   g. Move magic number to a cell
   h. Data clumps
   i. Shotgun surgery
   j. Data envy
   k. Keeping regular
   l. Convert multiple relative references to absolute
   m. Ways to suppress #DIV/0!
   n. Access external linked data safely
   o. Use the right kind of LOOKUP
   p. Check for legacy issues
2. **For data**
   a. Clean-up
   b. Blank or missing?
3. **For VBA Code**
   a. Extract method
   b. Comments
   c. Primitive obsession
   d. Replace array with object
   e. Replace magic number with symbolic constant
   f. Consolidate conditional expression
   g. Remove control flag
   h. Use Guard Clauses
   i. Replace nested conditional with guard clauses
   j. Introduce assertion
   k. Replace Error Code with Exception
   l. Replace Exception with Test
   m. VBE Tools-Options
   n. Speed Optimisation

## For worksheets

### Impose Style conventions

Use styles, fonts and colour with consistent meanings. Rather than only using one attribute, use the Style dialog to define styles such as "Parameters", "MonthlyInputs", "AnnualReport". If you decide to change the style for input cells, for example to make the font Courier or to put a border around each cell, you only have to modify this named style and all cells with that style show the modified format.

This only works if the conventions are followed. If you use blue text to mean an input cell, and someone else just thinks that blue is a nice colour, you are in the position described by Roedy Green [5] "Developers are lulled into complacency by conventions. By every once in a while, by subtly violating convention, you force him to read every [formula] with a magnifying glass"

Related to this are conventions of good practice such as print settings, sheet protection, removing circular references and replacing intentional ones by Goal Seek, adding cross-check formulas on each sheet that are echoed on a cover sheet, conditional formatting to highlight exceptions, and the use of buttons to run macros. If the spreadsheet contains any "Worst Practices"[10], fix them. There is debate on what constitute Best Practices [11,12,13] which are aired at every EuSpRIG conference[26]

### Follow Industry Conventions

In financial modelling, there are some specific guidelines that are recommended in that area. For example, on the level of documentation that is required for specific classes of users, unit denominations, that time periodicity should be consistent, and so on. Following these will considerably reduce the maintenance burden.

### Make it easy to see non-obvious content

Not all formulas are in cells; some can be in defined names, some in data validation, conditional formatting, chart series, and so on. Provide a macro [6] that can quickly produce a table of contents with hyperlinked sheet names, lists of defined names, distinct formulas in each sheet, formulas in other places, and so on. Then you can easily compare the current state with the previously documented state to discover what has changed.

### Duplicated code

The simplest duplicated code problem is when you have the same expression in two related formulas. Then all you have to do is Extract Expression and refer to the intermediate result from both places.
Example 1: replace multiple LOOKUP functions on the same range with one MATCH and multiple INDEX functions. This means there is only one time-consuming lookup and all the rest are the fast index function. It also makes the found row or column number explicit which may assist in understanding.
Example 2: this formula suppresses the #N/A! (Not Available) error from a lookup function:

B2=IF(ISNA(VLOOKUP(D2,$J$1:$M$10,2,FALSE)),"",VLOOKUP(D2,$J$1:$M$10,2,FALSE))

The VLOOKUP function is duplicated. Instead, place the lookup in its own column and refer to that:

B2=VLOOKUP(D2,$J$1:$M$10,2,FALSE)

C2=IF(ISNA(B2),"",B2)

Intermediate columns can be hidden or grouped if you don't need to see the results of the lookup. However, it is often better to know when the lookup has failed, so being able to see the results can be informative.

These intermediate cells are sometimes termed 'helper' columns or rows.
You could also use the IFERROR formula in Excel 2007 (or a UDF for earlier versions)
C4 =IFERROR(VLOOKUP(C3,FruitPrices,2,FALSE),0)

**Overlong formula**

*I would have written a shorter letter, but I did not have the time.* (Blaise Pascal)

Raffensperger[7] says "It is true that longer formulas are less readable, but the cost of a shorter formula can be more cells and a longer precedence tree, which may be less readable." Since the early days of programming people have realized that the longer a procedure is, the more difficult it is to understand; the same applies to formulas.

Developers new to fine-structured financial models may be surprised at the number of intermediate steps. When you have lived with such a spreadsheet for a few years, however, you learn just how valuable all those cells are in following the calculation chain of the model.

Example 3: Using line breaks to reveal the organisation of a long formula:

```
=IF($S10<>"", MAX($S10, $Q10, DATE($S$4, 12, 31), 0, W$4-
YEAR($Q10), $Q24, 1), 0)+IF($T10<>"", MAX($T10, $Q10, DATE($T$4,
12, 31), 0, W$4-YEAR($Q10), $Q24, 1), 0)+IF($U10<>"", MAX($U10,
$Q10, DATE($U$4, 12, 31), 0, W$4-YEAR($Q10), $Q24, 1),
0)+IF($V10<>"", MAX($V10, $Q10, DATE($V$4, 12, 31), 0, W$4-
YEAR($Q10), $Q24, 1), 0)+IF($W10<>"", MAX($W10, $Q10, DATE($W$4,
12, 31), 0, W$4-YEAR($Q10), $Q24, 1), 0)
```

Use Alt+Enter to create line breaks:

```
=IF($S10<>"",MAX($S10,$Q10,DATE($S$4,12,31),0,W$4-YEAR($Q10),$Q24,1),0)

+IF($T10<>"",MAX($T10,$Q10,DATE($T$4,12,31),0,W$4-YEAR($Q10),$Q24,1),0)

+IF($U10<>"",MAX($U10,$Q10,DATE($U$4,12,31),0,W$4-YEAR($Q10),$Q24,1),0)

+IF($V10<>"",MAX($V10,$Q10,DATE($V$4,12,31),0,W$4-YEAR($Q10),$Q24,1),0)

+IF($W10<>"",MAX($W10,$Q10,DATE($W$4,12,31),0,W$4-YEAR($Q10),$Q24,1),0)
```

Now we can see clearly, it is evident that:
   a) the max of 0 and 1 should be already known
   b) The constant term can be extracted eg Z10=MAX(Q10,$W$4-YEAR( Q10),Q24,1)
   c) The false IF value of 0 can be moved to the front to avoid confusion with the MAX
   d) The individual expressions that refer relatively to columns S..W can be entered as relative formulas eg AA10=IF(S10="",0,MAX($Z10,S10,DATE( S$4,12,31))), copy thru AE10.
   e) So the final formula is AF10=SUM(AA10:AE10); a total of six cells.

f) Or as an array formula without the intermediate columns Z10:AE10
   {=SUM(IF(S10:W10="", 0, MAX(1,Q10,$W$4-YEAR(
   Q10),Q24,S10:W10,DATE(S$4:W$4,12,31))))}

### Simplify formulas

A good worker takes the time to learn their craft. Following the discussions on the Excel forums and mail lists will expose the average Excel user to more efficient ways of implementing a given calculation. For example, the SUMPRODUCT function sums the product of corresponding cells in two ranges, giving the same result as the sum of a third range of multiplication formulas. If the only figure of interest is the total, then SUMPRODUCT is a simpler solution.

There is a risk that one who uses more sophisticated functions may have to hand over the spreadsheet to another not familiar with these features who may break them. That is a training issue, and it is generally recognised that a common problem is the self-taught nature of most Excel users who have never taken a course or obtained a qualification in the tool they use every day.

Another way to hide complexity from the worksheet is to move awkward formulas to VBA User Defined Functions (UDFs). The cost is a slower performance; the benefit is making the spreadsheet more understandable at a surface level.

## Move magic number to a cell

Remove a numeric constant ('magic number') from a formula and isolate it in a cell of its own. For further clarity, give that cell a name with workbook scope.
A figure, assumed to be constant at the time the spreadsheet was created, may turn out in time to need revision, and therefore it is really a variable. An example is a tax rate or a conversion factor. One commonly overlooked is a column number in a VLOOKUP function; the data may start off in column 3 but it is easy to restructure the table and forget that dependency. To be safer, use a MATCH function that finds a unique column heading in the table and returns the column number to be used. Excel 2007 has more flexible structured data table references.
The benefit of putting a constant into its own cell is that if its value ever does change, it need only be changed in one place. If it was a constant repeated in several formulas, some instances may not be changed, which leads to inconsistent calculations and incorrect results. You don't need to do this to numbers that really are constant like 1 or 100 used in percentage calculations.

### Data clumps

Bunches of data that hang around together really ought to be made into their own named range. Often you see a particular list of data items that tend to be passed together. Several formulas may use this list, either on one sheet or in several sheets. Such a group of data is a data clump and can be replaced with a range name that carries all of this data.
A good test is to consider deleting one of the data values: if you did this, would the others make any sense? For example, a start and end date range belong together.

### Shotgun surgery

Shotgun surgery is when every time you make a kind of change, you have to make a lot of little changes to a lot of different worksheets. When the changes are all over the place, they are hard to find, and it's easy to miss an important change. Organize the flow so that all the changes are in the same place.

**Data envy**

The whole point of worksheets is that they are a technique to package data with the calculations used on that data. A classic 'bad smell' is a formula that seems more interested in a sheet other than the one it actually is in. The most common focus of the envy is the data, where a method pulls several values from another sheet to calculate some value. Fortunately the cure is obvious, the calculation clearly wants to be elsewhere, so you move it there and refer to that result cell.

**Keeping regular**

Each row in a financial model should contain just one formula repeated across. This prevents accidental overwriting of the 'special' parts later by copying across.

An inconsistency arises when one formula in a block of cells is different from those in its neighbourhood. Possible causes are:
1. A user is typing repeated formulas individually rather than creating copies (or clones) using either copy & paste or the drag or fill commands. Because of all the repeated typing, they forget some references in some formulas that they include in others.
2. A user types a constant into a cell in a mistaken attempt to correct a formula.
3. A formula omits adjacent nonempty cells.

Excel 2002 and later helps to some extent by flagging inconsistent cells in a block with a green triangle in the upper left corner.

**Convert multiple relative references to absolute**

A user who does not know absolute references can break the 'keeping regular' rule and create formulas like this:

|   | A | B | C | D | E |
|---|---|---|---|---|---|
| 1 | **Item** | **Gross** | **Net after** | 0.1 | **discount** |
| 2 | Apple | 10 | =B2*(1-D1) | | |
| 3 | Orange | 12 | =B3*(1-D1) | | |
| 4 | Banana | 9 | =B4*(1-D1) | | |
| 5 | Grape | 15 | =B5*(1-D1) | | |

This gives rise to many separate formulas, rather than one base formula that can be copied down. In order to fix this,
1. Create a name for cell D1, "DiscountRate".
2. In the Formulas ribbon, Defined Names Group, click Defined Name dropdown, click Apply Names, check "Ignore Relative/Absolute", uncheck "Use row and column names", click OK

VBA code:
```
Cells.ApplyNames Names:=" DiscountRate ",
IgnoreRelativeAbsolute:=True, _
    UseRowColumnNames:=False, OmitColumn:=True,
    OmitRow:=True, Order:=1, AppendLast:=False
```

**Ways to suppress #DIV/0!**

Usually you want to know when a number is being divided by zero. There may be cases when a division is setup beforehand for a year and the data only becomes available over time. In that case you can suppress the error value by
1. Using an IF test to show 0 instead =IF(C4=0,0,C3/C4)
2. Conditional formatting any error with the same font as background colour
3. Leaving the error on the screen but in Page Setup set the Sheet options "Print Cell errors as" to the symbol you wish to see on paper, such as a dash.

**Access external linked data safely**

In Excel, all data is global. Any user can dip into any cell in any worksheet of any workbook and pull in the data they want. Worse, the owner of the source spreadsheet has no idea that they are doing that. If the users refer to cells by address and the owner changes the structure, the data retrieved is invalid. One alternative is to lock down source workbooks and make them read-only. Therefore if the owner needs to update the source, they must make a copy ('fork the code') and hope that everyone who uses the data is aware of the change. If they simply rename or move the source workbook in the hope that that will alert users to missing source files, sadly those users who have the option to update links turned off, or who always click 'Don't update' because that's what they always did, will never know that the source is gone away, until the one day on financial period close when they suddenly discover it.

If your data is to be used by others, firstly protect it or place the source workbook in a folder to which they have not got rights. Secondly, marshal and publish all data that you intend to share to a known location that you undertake to be the defined interface to your data. Thirdly, ensure all the users use data queries that are refreshed upon open.

Replace multiple references to worksheets in an external workbook with a single reference to a consolidated figure. This is a special case of designing an interface, a data marshalling sheet to make it clearly visible where the source and use of external data is visible.

Apply a range name to each distinct block of cells in a source spreadsheet that you intend to link to from a target spreadsheet. Cell addresses can become invalid if the source workbook is changed. Range names will continue to refer to the same cell. When entering the external link in the target spreadsheet, refer to the address in the source spreadsheet by its range name rather than by its cell address, eg, C5.

This avoids the problem that arises where:
1. a formula in the target sheet refers to a source cell by address;
2. a user moves the location of the data in the source sheet;
3. the reference in the target sheet is not changed to correspond;
4. as a result data is imported from the wrong location.

Don't do: ='F:\Topdir1\[source1.xls]Sheet1'!$A$2
Do this: ='F:\Topdir1\[source1.xls]Sheet1'!Somename
or this: ='F:\Topdir1\source1.xls'!Atest

To search in a workbook for cells containing links, search for the exclamation mark (!) in formulas. You should maintain documentation of these external links. Auditing tools can also list them.

**Use the right kind of LOOKUP**

It is quite common to find spreadsheets where the user has not provided the fourth argument to the VLOOKUP function and therefore does not realise that it may be returning incorrect values. The syntax is:

VLOOKUP(lookup_value,table_range,col_index_num,range_lookup)
The fourth argument range_lookup is an optional logical value that specifies whether you want VLOOKUP to find an exact match or an approximate match. If range_lookup is omitted or TRUE, an approximate match is returned. If the first value in the table is greater than lookup_value, VLOOKUP returns #N/A, otherwise it finds the last row where the value is equal to or not greater than lookup_value.
If range_lookup is FALSE, VLOOKUP matches the largest value in table_range that is less than or equal to lookup_value.
If you omit optional arguments, the program will assume default values. The default behaviour of a function may not be what you expect or want. It is better to explicitly specify option arguments in order to remove any ambiguity.

**Check for legacy issues**

Some old workbooks may still have Lotus options set. This can give rise to lookup problems described in Microsoft Knowledgebase articles.
Some problems with Transition Formula Evaluation (TFE) or Alternate Expression Evaluation (AEE) are detailed on the Microsoft support site.
http://support.microsoft.com/default.aspx?scid=kb;en-us;87442
VLOOKUP()/HLOOKUP() Return Incorrect Value with TFE or AEE
http://support.microsoft.com/default.aspx?scid=kb;en-us;94202

There could be many other issues in legacy workbooks such as numeric values being stored as text, or misleading pie charts being used instead of column charts. However, refactoring is dealing with the improvement of working models rather than fixing errors, so these are not covered here.

## *Data refactoring*

Data refactoring[16] is a small change to its organisation which improves its design without changing its semantics. In Excel, tables are usually non-normalised and partly pivoted which can create considerable work to convert them to database structures.

Structural. A change to the table structure of your data. Adding calculated columns may permit the use of pivot tables rather than doing calculations on matrix layouts. Simon Murphy says "I can't remember the last time I wrote a big complex mainly formula based reporting/analysis app. As soon as the client wants multiple views (say monthly and annual) I am thinking big table driven pivots."

Data quality. A change which improves and/or ensures the consistency and usage of the values stored within the database. See Arxiv.org [20] for this author's articles on Data Quality in spreadsheets. A common spreadsheet task is to clean up inconsistent data[18,19]. For example, you might need to ensure that all phone numbers are in a common format, email addresses are in lowercase, dates are formatted consistently, and so on. John Walkenbach's Power Utility Pack [17] add-in contains a few useful tools for data clean-up.

Referential integrity. A change which ensures that a referenced row exists within another table and/or that ensures that a row which is no longer needed is removed appropriately. This is particularly onerous in a spreadsheet because you have to do this by manual operations.

**Don't leave input cells blank**

If you do, you can not then distinguish between missing data and empty data. Possible refactorings are:
1. Enter a zero. Use the workbook Options if you wish to hide zero values.
2. If a valid result requires every cell to have an entry, pre-fill the cells with =NA() to mean "Not Available". This error value will propagate to all cells that depend on the input cell and so if the final result shows #N/A! you cannot miss the fact that some input has not been supplied.

## *Refactoring VBA*

We are on more familiar refactoring territory here – after all VBA is code. However its limitations make many of the standard object-oriented refactorings not applicable. The following list is quite small compared to the VB patterns [8,9] for which there is tool support such as CodeShine, Aivosto, and Refactor! for VB.Net. For VBA one can use Aivosto[23], SmartIndenter[22], MZTools[24] and VBLiteUnit[25] for Unit Testing

**Extract method**

Split procedures into logical separate procedures. For example, an all-in-one routine that gets input from the user and then embarks on a large process could be divided into two. The processing routine that receives its parameters passed in could now be called from another context, permitting code reuse or automation or unit testing.

Don Roberts's guideline is "The first time you do something, you just do it. The second time you do something similar, you wince at the duplication, but you do the duplicate thing anyway. The third time you do something similar, you refactor."

**Comments**

A heuristic Kent Beck recommends is that whenever we feel the need to comment something, we write a method instead. Such a method contains the code that was commented but is named after the intention of the code rather than how it does it.
The key here is not method length but the semantic distance between what the method does and how it does it. There's a fine line between comments that illuminate and comments that obscure. Are the comments necessary? Do they explain "why" and not "what"?

**Replace array with object**

You have an array in which certain elements mean different things. Replace the array with an object that has a field for each element.
Arrays are a common structure for organizing data. However, they should be used only to contain a collection of similar objects in some order. Sometimes, however, you see them used to contain a number of different things. Conventions such as "the first element on the array is the person's name" are hard to remember. With an object you can use names of fields and methods to convey this information so you don't have to remember it or hope the comments are up to date.

Collections and Scripting.Dictionary objects can be more convenient than arrays. Take advantage of the uniqueness of collection objects and the ease of addressing and update of Dictionary objects.

Excel's implementation of user defined data types has limitations. Some authors [27] recommend creating classes, for example to ensure that settings are reset when an object goes out of scope.

### Replace magic number with symbolic constant

You have a literal number with a particular meaning. Create a constant, name it after the meaning, and replace the number with it. This can be a half-way step to "Replace Array with Object" if you replace references to aPerson(1) with aPerson(PERSON_NAME). If you restructure the array you only have to change the symbolic constants rather than look for all index references to the array elements.

Use named ranges rather than fixed cell addresses when referring to worksheet data in code.

### Consolidate conditional expression

Sometimes you see a series of conditional checks in which each check is different yet the resulting action is the same. When you see this, you should use ands and ors to consolidate them into a single conditional check with a single result.

### Remove control flag

You have a variable that is acting as a control flag for a series of Boolean expressions. Use a break or return instead. When you have a series of conditional expressions, you often see a control flag used to determine when to stop looking. Such control flags are more trouble than they are worth.

### Use Guard Clauses

A Guard Clause is a chunk of code at the top of a function (or block) that serves a similar purpose to a Precondition. It typically does one (or any or all) of the following:

- checks the passed-in parameters, and returns with an error if they're not suitable.
- checks the state of the object, and bails out if the function call is inappropriate.
- checks for trivial cases, and gets rid of them quickly.

### Replace nested conditional with guard clauses

The key point about Replace Nested Conditional with Guard Clauses is one of emphasis. If you are using an if-then-else construct you are giving equal weight to the If leg and the Else leg. This communicates to the reader that the legs are equally likely and important. Instead the guard clause says, "This is rare, and if it happens, do something and get out."

### Introduce assertion

A section of code assumes something about the state of the program. Make the assumption explicit with an assertion. Such assumptions often are not stated but can only be decoded by looking through an algorithm. Sometimes the assumptions are stated with a comment. A better technique is to make the assumption explicit by writing an assertion:
```
Debug.Assert Len(cell.value) > 0
```

### Replace Error Code with Exception

A method returns a special code to indicate an error. Throw an exception instead.

The problem is that the part of a program that spots an error isn't always the part that can figure out what to do about it. When such a routine finds an error, it needs to let its caller know, and the caller may pass the error up the chain. Java has a better way: exceptions. Exceptions are better because they clearly separate normal processing from error processing. This makes programs easier to understand, and as I hope you now believe, understandability is next to godliness

### Replace Exception with Test

You are throwing a checked exception on a condition the caller could have checked first. Change the caller to make the test first.

### VBE Tools-Options

See Dave Hawley's Excel/VBA Golden Rules[14]

- Switch off *Auto Syntax Check* on the Editor tab. You don't want message boxes when you type a syntax error - you just want the problem line shown in red so you can fix it later.

- Switch on *Require Variable Declaration* on the Editor tab. This automatically inserts *Option Explicit* at the top of all modules.

- Declare the variable type when you dimension a variable; *Option Explicit* does not require this, but you should do it.

- Switch off *Compile On Demand* on the General tab. You want to be told about syntax errors immediately when you run your code, not just when the dodgy routine actually gets called.

- Set Error Trapping to *Break on unhandled errors* on the General tab.

- Prompt to save changes when program starts, on the Environment tab.

- Enter code in lower case, so you can instantly see a typo because the variable name does not become capitalised when you press enter.

- While it is a good objective to keep all related procedures in the same module, Excel has a 64KB module size limit that may make that impractical.

### Speed Optimisation

Michael Jackson: "The First Rule of Program Optimization: Don't do it. The Second Rule of Program Optimization (for experts only!): Don't do it yet." The secret to fast software, in all but hard real-time contexts, is to write tuneable software first and then to tune it for sufficient speed.  Use time measurement addins such as FastExcel[15] by Charles Williams to find that small part of the spreadsheet where the performance sinks.

Examine code that loops through cells and look for faster alternatives like the Range.Find method, Autofilter, Advancedfilter, and SpecialCells..

You almost never need to select or activate an object to change properties or call a method. If you must select a range object, use Application.goto.

Turn off Calculation and ScreenUpdating and Events for slow routines. Assume your macro will fail, so turn back on in the exit point of the routine.

Avoid writing data back to the spreadsheet cell by cell. Accumulate data in blocks of arrays and write the array back in one operation for speed.

The With statement can make code easier to read when performing several different operations on the same object.

Iif(Boolean, trueExpression,FalseExpression) is slower than If-Else blocks because both expressions are always evaluated. If you intend both to be evaluated, it is clearer to explicitly do that in code and store the return values for the IIF function. Some languages do short-cut evaluation, so in case the code is to be read by someone who knows more than VBA, avoid a possible misreading of the code.

http://www.databaserefactoring.com/